\documentclass[letterpaper]{article} 
\usepackage[preprint]{aaai2027} 
\usepackage[hyphens]{url} 
\usepackage{graphicx} 
\usepackage{natbib} 
\usepackage{caption} 
\usepackage{amsmath}
\usepackage{amsfonts}
\usepackage{booktabs}
\usepackage{pifont}

\title{Beyond Reconstruction: Full-Context Generative DiT for Music Generation}
\author{\fontsize{11}{14}\selectfont\mdseries
Yunjia Li\textsuperscript{\rm 1,\rm 2,\textdagger},
Menglin Wu\textsuperscript{\rm 2,\textdagger},
Junyu Dai\textsuperscript{\rm 2,\textdagger},
Xinyue Fan\textsuperscript{\rm 2,\textdagger},
Xiangang Li\textsuperscript{\rm 2,\textdagger},\\
Haoxu Wang\textsuperscript{\rm 2,\textdagger},
Jianwei Yu\textsuperscript{\rm 2,\textdagger},
Huaicheng Zhang\textsuperscript{\rm 2,\textdagger},
Han Zhao\textsuperscript{\rm 2,\textdagger},
Weiqin Li\textsuperscript{\rm 2,\textdaggerdbl},\\
Yufei Shi\textsuperscript{\rm 2,\textdaggerdbl},
Cheng Wen\textsuperscript{\rm 2,\textdaggerdbl},
Sitong Zhao\textsuperscript{\rm 2,\textdaggerdbl},
Qixi Zheng\textsuperscript{\rm 2,\textdaggerdbl},
Haina Zhu\textsuperscript{\rm 2,\textdaggerdbl},
Wei Li\textsuperscript{\rm 1,\textdaggerdbl}
}
\affiliations{
\textsuperscript{\rm 1}College of Computer Science and Artificial Intelligence, Fudan University\\
\textsuperscript{\rm 2}Alibaba Token Foundry
}

\begin{document}

\maketitle

\begingroup
\renewcommand{\thefootnote}{\fnsymbol{footnote}}
\footnotetext[2]{Core authors.\qquad \textsuperscript{\textdaggerdbl}\,Contributors.}
\endgroup

\begin{abstract}
Hybrid music generators combine the long-range planning of an autoregressive
language model with the fidelity of a diffusion- or flow-based acoustic
renderer. Yet renderers are trained with clean, target-derived codec tokens but
deployed with imperfect language-model predictions, creating
\emph{codec-interface exposure bias}. Rather than treating rendering as a
simple reconstruction task, we formulate it as full-context
generation from an imperfect discrete plan. We introduce FullDiT, a conditional
DiT that fuses eight frame-aligned RVQ streams with independently encoded
captions and lyrics and uses non-causal self-attention over the complete
acoustic latent sequence. During training, Error-Matched Distractor
Conditioning (EMDC) matches per-codebook replacement rates to teacher-forced
top-1 error rates and samples near-miss tokens from cosine-KNN neighborhoods
without changing the acoustic target. At inference, four-way classifier-free
guidance (4-CFG) independently scales codec, lyric, and caption guidance
increments. Matched ablations show that EMDC improves ViSQOL by 0.77 under
synthetic corruption and is clearly preferred in non-tied comparisons with
fixed language-model tokens. Further ablations show gains from full-song
context and renderer-side text conditioning. The complete system outperforms
five commercial systems on 15 of 18 automatic metrics and ranks among the top
three on the Artificial Analysis Music with Vocals Leaderboard. The demo page
is available at \url{https://selinacloudl.github.io/fulldit-demo/}.
\end{abstract}


\section{Introduction}
\label{sec:introduction}

Music generation must coordinate long-range form, lyrics, and instrumentation
with fine-grained acoustics over several minutes. Existing systems use direct
diffusion or flow generation
\citep{huang2023noise2music,evans2024stableaudio,ning2025diffrhythm,gong2025acestep},
autoregressive codec modeling
\citep{dhariwal2020jukebox,agostinelli2023musiclm,copet2023musicgen}, or hybrid
pipelines that separate discrete planning from generative acoustic rendering
\citep{lam2023melody,bai2024seedmusic,zhang2025inspiremusic,yang2025songbloom}.
This hybrid design can combine long-range musical planning with high-fidelity
acoustic realization, but it creates an interface shift: renderers are
typically trained with clean codec tokens extracted from their acoustic targets
but deployed with imperfect language-model predictions. Unrealistically clean
training conditions can make a generative renderer behave like a paired
reconstruction decoder and propagate token errors at inference. We call this
shift \emph{codec-interface exposure bias}.

Uniform corruption is an incomplete remedy because prediction difficulty
varies across RVQ codebooks and incorrect tokens can remain locally plausible.
The renderer must also reconcile frame-aligned codec tokens, sequential lyrics,
global captions, and full-song dependencies, deciding when to trust the plan
and when to complete it from text, context, and its acoustic prior.

We formulate rendering as full-context generation from an imperfect
discrete plan. Architecturally, FullDiT fuses eight frame-aligned RVQ streams
with independently encoded captions and lyrics, and applies non-causal
self-attention over the complete acoustic latent sequence. During training,
Error-Matched Distractor Conditioning (EMDC) matches each codebook's
replacement rate to its teacher-forced top-1 error rate and samples near-miss
tokens from cosine-KNN neighborhoods while leaving the acoustic target
unchanged. At inference, four-way classifier-free guidance (4-CFG)
independently scales codec, lyric, and caption increments.

Our contributions are:
\begin{itemize}
    \item We introduce a full-context conditional generative acoustic DiT that
    combines frame-aligned multi-codebook RVQ conditions, renderer-side caption
    and lyric conditioning, and non-causal full-song context in a continuous
    flow model. At inference time, 4-CFG independently scales codec, lyric, and
    caption increments.
    \item We propose EMDC, an error-rate-matched, geometry-guided corruption
    scheme that exposes the renderer to realistic codec-token errors during
    training while preserving the acoustic target.
    \item We validate FullDiT under clean, synthetically corrupted, and
    LM-generated codec conditions. Matched ablations isolate EMDC, full-song
    context, and renderer-side text conditioning; additional studies cover a
    4-CFG sweep, comparisons with five commercial systems, and an external
    public blind test. EMDC improves ViSQOL by 0.77 under synthetic token
    corruption and is
    clearly preferred in blind listening with fixed language-model tokens; the
    complete system outperforms five commercial systems on 15 of 18 automatic
    quality metrics and ranks in the top three on the Artificial Analysis Music
    with Vocals Leaderboard.
\end{itemize}

\section{Related Work}
\label{sec:related}

\subsection{Three Paradigms of Music Generation}

\paragraph{Direct diffusion and flow generation.}
Noise2Music, Stable Audio, DiffRhythm, and ACE-Step generate spectrograms,
waveforms, or continuous latents with diffusion or flow models
\citep{huang2023noise2music,evans2024stableaudio,ning2025diffrhythm,gong2025acestep,peebles2023dit}.
They avoid a discrete planning interface but must jointly represent long-range
structure, text alignment, and local acoustics in one generator.

\paragraph{Autoregressive codec-token generation.}
SoundStream and EnCodec establish the encoder--RVQ--decoder pattern
\citep{zeghidour2022soundstream,defossez2023encodec}, while Jukebox, MusicLM,
and MusicGen model multi-scale, hierarchical, or interleaved music tokens
\citep{dhariwal2020jukebox,agostinelli2023musiclm,copet2023musicgen}.
Low-frame-rate discretization shortens the planning sequence, but quality still
depends on both token prediction and acoustic decoding.

\paragraph{Hybrid token planning and generative rendering.}
Hybrid systems couple an autoregressive language model for discrete planning
with a diffusion- or flow-based renderer for high-fidelity acoustic
realization. MeLoDy, Seed-Music, InspireMusic, SongBloom, and HeartMuLa
instantiate this pattern with different intermediate representations and
coupling strategies
\citep{lam2023melody,bai2024seedmusic,zhang2025inspiremusic,yang2025songbloom,yang2026heartmula}.
This decomposition makes the planning--rendering interface explicit. When a
renderer is trained with target-derived codec tokens but deployed with planner
predictions, the interface induces the codec-interface exposure bias studied
here.

\subsection{Full-Context Conditional Acoustic Rendering}

The key renderer-side distinction is which conditions are available and over
what temporal span. In Seed-Music's audio-token pipeline, multimodal controls
condition the autoregressive generator, while the diffusion renderer receives
the generated audio tokens \citep{bai2024seedmusic}. InspireMusic likewise
sends text to the autoregressive model and conditions its super-resolution
flow renderer on a single coarse token stream
\citep{zhang2025inspiremusic}. SongBloom operates patch-wise: its diffusion
stage receives current-patch sketch tokens, an autoregressive hidden vector
that can carry text semantics, and acoustic latents from the preceding patch
\citep{yang2025songbloom}. HeartMuLa conditions its language model on style,
lyrics, and reference audio, whereas HeartCodec's flow decoder is described as
receiving low-rate quantized features rather than separately encoded text
\citep{yang2026heartmula}. Across these hybrid pipelines, no renderer is
described as jointly consuming a frame-aligned multi-codebook plan,
independently encoded captions and lyrics, and the complete predicted codec
sequence through non-causal attention.

Closely related renderer-level work studies generative codec resynthesis.
Multi-Band Diffusion synthesizes high-fidelity audio from low-bitrate discrete
representations, while Liu et al. study speech resynthesis from coarse codec
tokens and show that both the learning target and the generative resynthesis
design materially affect output quality
\citep{sanroman2023multibanddiffusion,liu2024codecresynthesis}. These results
establish that a generative renderer can recover acoustic detail beyond a
deterministic codec decoder. FullDiT advances this direction from codec
resynthesis to full-song conditional generation from a language-model-predicted
multi-codebook plan, with renderer-side text conditioning and bidirectional
song context.

\subsection{Robust Conditioning from Imperfect Upstream Outputs}

Imperfect discrete conditioning can produce audible artifacts
\citep{sanroman2023multibanddiffusion}, motivating training procedures that
expose a downstream renderer to realistic deviations from oracle codec tokens.
Related techniques address different forms of condition mismatch. Scheduled
sampling reduces exposure bias in autoregressive sequence models by
occasionally feeding model-generated previous tokens instead of ground-truth
tokens during training \citep{bengio2015scheduledsampling}. Like EMDC, it
exposes a model to imperfect conditions during training, but it perturbs the
model's own autoregressive history rather than an external codec plan.
Conditioning augmentation in cascaded image diffusion perturbs an upstream
continuous output with Gaussian noise or blur
\citep{ho2022cascaded}. Condition dropout jointly trains conditional and
unconditional predictions for classifier-free guidance
\citep{ho2022cfg}. These techniques motivate perturbing or dropping training
conditions, but do not specify how to corrupt a multi-codebook codec plan
according to the error profile of a frozen music language model. EMDC fills
this gap by matching replacement frequency to teacher-forced error rates for
each codebook and sampling plausible error tokens from cosine-KNN
neighborhoods in the corresponding codec space.

\section{Method}
\label{sec:method}

\begin{figure*}[!t]
    \centering
    \includegraphics[
        width=\textwidth,
        keepaspectratio
    ]{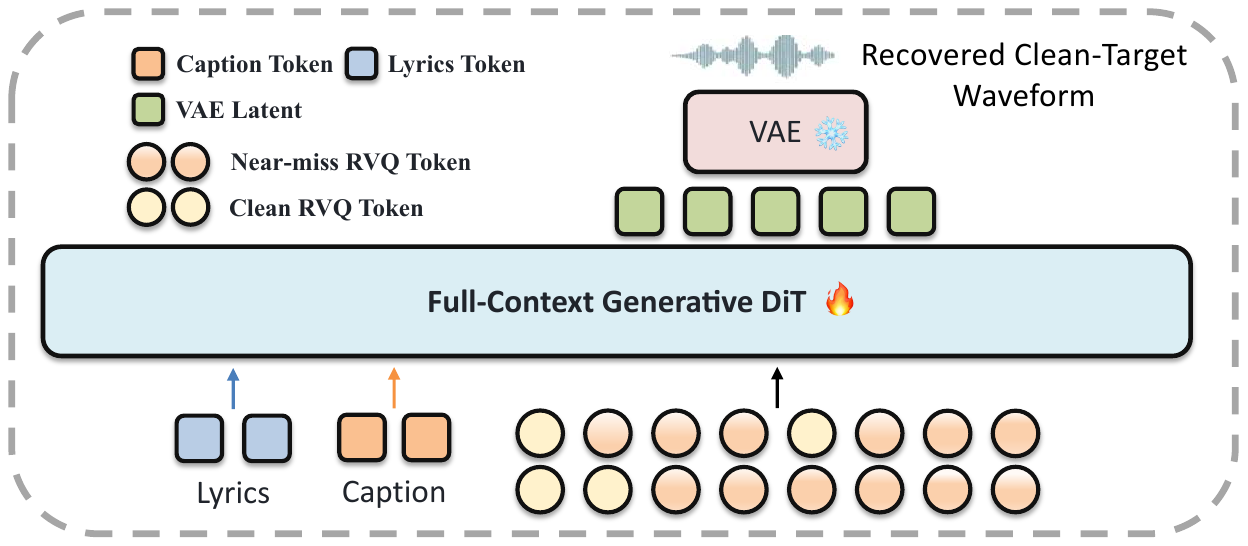}
    \caption{FullDiT acoustic renderer. Independently encoded lyrics and
    captions provide textual conditions, while frame-aligned multi-codebook
    RVQ tokens provide the discrete song plan. The illustrated plan mixes
    clean tokens with plausible near-miss substitutions. EMDC constructs this
    imperfect condition to approximate errors in a language-model-predicted
    plan, but changes only the codec condition: the paired clean target
    waveform and its target VAE latent remain unchanged. Rather than
    reconstructing the corrupted token plan, FullDiT uses text and non-causal
    full-song context to generatively recover a complete VAE-latent sequence
    consistent with the unchanged clean target. A frozen VAE decoder then maps
    the recovered latents to waveform.}
    \label{fig:overview}
\end{figure*}

\subsection{System Overview}
\label{sec:overview}

Given user-provided lyrics and a caption, the complete music generation system
comprises a semantic-aware RVQ tokenizer,
a hierarchical autoregressive language model, an optional two-level melody
module, and FullDiT. The tokenizer maps raw music to eight RVQ token streams
and uses Conformer-based \citep{gulati2020conformer} BEST-RQ-style pretraining
\citep{chiu2022bestrq}, multi-task fine-tuning, and discrete-token training to
preserve semantic content and reconstruction fidelity. The hierarchical
language model pairs an
8B global LLM, which autoregressively predicts the first-level stream from
lyrics, captions, and musical attributes, with a 0.4B local LLM that predicts
the residual streams, supporting both vocal and instrumental generation. For
cover generation, the melody module discretizes note- and frame-level pitch
cues into coarse and fine tokens that control the global vocal melody and local
pitch detail. This work focuses on FullDiT, an 8B-parameter non-causal DiT that
performs full-song flow matching in a continuous VAE latent space, conditioned
on the complete codec sequence, caption, and lyrics; a frozen VAE
decoder maps its output to 48-kHz stereo audio. Accordingly,
Figure~\ref{fig:overview} shows only FullDiT and its conditioning interface.
Controlled studies use separately trained, matched 1.5B variants.

\subsection{Full-Context Conditional Acoustic Rendering}
\label{sec:fullcontext}

For a paired training example
$(x,y_{\mathrm{cap}},y_{\mathrm{lyr}})$, the tokenizer extracts codec tokens
$c^{\mathrm{gt}}$, while a fixed VAE encoder produces the target acoustic
latent $z_0$:
\begin{equation}
\begin{array}{@{}l@{\;}c@{\;}l@{}}
c^{\mathrm{gt}} & = & Q_{\mathrm{RVQ}}(x),\\
z_0 & = & E_{\mathrm{VAE}}(x),\\
\widetilde{c} & \sim & q_{\mathrm{EMDC}}
  (\cdot\mid c^{\mathrm{gt}}).
\end{array}
\label{eq:training-pair}
\end{equation}
EMDC perturbs only the codec condition and leaves the target latent $z_0$
unchanged. FullDiT models
$p_\theta(z_0\mid\widetilde{c},y_{\mathrm{cap}},y_{\mathrm{lyr}})$.
At inference, the language-model-generated plan
$\widehat{c}\sim
p_{\mathrm{LM}}(\cdot\mid y_{\mathrm{cap}},y_{\mathrm{lyr}})$
replaces $\widetilde{c}$ as the renderer input.

\subsubsection{Frame-Aligned Codec Conditioning and Full-Song Context}

Let $c_{i,k}$ be the token at frame $i$ in codebook $k$. FullDiT performs a
codebook-specific lookup and injects the fused embedding into the matching
audio frame:
\begin{equation}
\begin{array}{@{}l@{\;}c@{\;}l@{}}
e_i^{\mathrm{codec}} & = & \mathrm{Fuse}_{k=0}^{7} E_k(c_{i,k}),\\
a_i^{(0)} & = & E_{\mathrm{audio}}(z_{t,i}) + e_i^{\mathrm{codec}}.
\end{array}
\label{eq:codec-injection}
\end{equation}
Frame-wise addition preserves the temporal correspondence between the
discrete plan and the acoustic latent. The eight codebook embeddings are
summed and projected to the audio hidden dimension. FullDiT then applies
non-causal self-attention over the complete acoustic sequence. Each frame can
therefore combine its aligned codec condition with both past and future
acoustic states, allowing a repeated chorus, a later motif, or a distant
section to inform local realization.

\subsubsection{Renderer-Side Caption and Lyric Conditioning}

Although captions and lyrics are already provided to the hierarchical
autoregressive language model, relying on its predicted codec plan as the sole
interface creates an information bottleneck: textual details may be omitted or
attenuated during plan generation. We therefore provide the original captions
and lyrics directly to FullDiT as renderer-side conditions, allowing the
acoustic renderer to consult the source text directly rather than relying
entirely on the predicted codec plan to convey all conditioning information. At every
block, audio states form the queries, and lyric, caption, and audio states form
the key--value context:
\begin{equation}
\begin{array}{@{}l@{\;}c@{\;}l@{}}
h_{\mathrm{cap}} & = & E_{\mathrm{cap}}(y_{\mathrm{cap}}),\\
h_{\mathrm{lyr}} & = & E_{\mathrm{lyr}}(y_{\mathrm{lyr}}),\\
Q & = & W_Q a,\\
K & = & [W_K h_{\mathrm{lyr}};W_K h_{\mathrm{cap}};W_K a],\\
V & = & [W_V h_{\mathrm{lyr}};W_V h_{\mathrm{cap}};W_V a].
\end{array}
\label{eq:text-context}
\end{equation}
Captions provide genre, instrumentation, mood, and production semantics;
lyrics provide phonetic content, vocal cues, and section structure.

\subsubsection{Flow-Matching Objective}

Following flow matching \citep{lipman2023flowmatching}, let
$\sigma\in[0,1]$ denote the path coordinate, with $\sigma=0$ corresponding to
data and $\sigma=1$ to Gaussian noise. Given a target latent $z_0$, we define
the linear conditional path and its target velocity as
\begin{equation}
\begin{array}{@{}l@{\;}c@{\;}l@{}}
\epsilon & \sim & \mathcal{N}(0,I),\\
z_\sigma & = & (1-\sigma)z_0+\sigma\epsilon,\\
v_\sigma^\star
  & = & \displaystyle\frac{\partial z_\sigma}{\partial \sigma}
  = \epsilon-z_0.
\end{array}
\label{eq:flow-path}
\end{equation}
Let $y=(y_{\mathrm{cap}},y_{\mathrm{lyr}})$. The model regresses the
conditional vector field with
\begin{equation}
\mathcal{L}_{\mathrm{FM}} =
\mathbb{E}_{z_0,\widetilde{c},y,\epsilon,\sigma}\!\left[
\left\|
v_\theta(z_\sigma,\sigma,\widetilde{c},y)-v_\sigma^\star
\right\|_2^2
\right].
\label{eq:flow-loss}
\end{equation}
At inference, sampling starts from $z_1\sim\mathcal{N}(0,I)$ and integrates
the learned field backward from $\sigma=1$ to $\sigma=0$. EMDC changes only
the codec input $\widetilde{c}$; it never changes $z_0$ or
$v_\sigma^\star$. A replacement
token therefore does not become a new acoustic label.

\subsection{Error-Matched Distractor Conditioning}
\label{sec:emdc}

We apply EMDC during training to construct a corrupted codec condition
$\widetilde{c}$ from the clean sequence $c^{\mathrm{gt}}$ while keeping the
target latent $z_0$ unchanged. For each frame $i$ and codebook
$k\in\{0,\ldots,7\}$, we
independently sample a replacement mask:
\begin{equation}
M_{i,k}\sim\mathrm{Bernoulli}(p_k),\qquad
p_k=1-\mathrm{Acc@1}_k.
\label{eq:emdc-mask}
\end{equation}
If $M_{i,k}=0$, the clean token is retained. Otherwise, a replacement $j$ is
sampled from the cosine top-$K_k$ neighborhood
$\mathcal{N}_k(c_{i,k})$, excluding the token itself:
\begin{equation}
q_k(j\mid c)=
\underset{j\in\mathcal{N}_k(c)}{\mathrm{softmax}}
\left(\cos(e_k(j),e_k(c))/\tau\right).
\label{eq:emdc-replacement}
\end{equation}

The teacher-forced top-1 error rate determines how frequently tokens in each
codebook are replaced. Cosine-KNN is an embedding-geometry-guided heuristic
for selecting the replacement token.

\begin{table}[t]
\centering
{
\small
\setlength{\tabcolsep}{1mm}
\begin{tabular*}{0.95\columnwidth}{@{\extracolsep{\fill}}cccccc@{}}
\toprule
\textbf{CB} & \textbf{Loss} & \textbf{PPL} & \textbf{Acc@1} &
$\boldsymbol{p_k}$ & $\boldsymbol{K_k}$ \\
\midrule
0 & 2.27 & 10 & 0.375 & 0.625 & 20 \\
1 & 2.26 & 10 & 0.379 & 0.621 & 20 \\
2 & 2.51 & 12 & 0.340 & 0.660 & 20 \\
3 & 2.82 & 17 & 0.302 & 0.698 & 30 \\
4 & 3.25 & 26 & 0.230 & 0.770 & 50 \\
5 & 3.65 & 39 & 0.189 & 0.811 & 80 \\
6 & 4.13 & 62 & 0.141 & 0.859 & 125 \\
7 & 4.49 & 89 & 0.108 & 0.892 & 180 \\
\bottomrule
\end{tabular*}
}
\caption{Codebook-wise EMDC calibration. Perplexity is approximated by
$\exp(\mathrm{loss})$. $K_k$ is a heuristic neighborhood width.}
\label{tab:emdc}
\end{table}

As Table~\ref{tab:emdc} shows, deeper residual codebooks are substantially
harder for the upstream predictor. A uniform replacement rate would therefore
understate their uncertainty. We choose $K_k$ by rounding approximately
$2\exp(\mathrm{loss}_k)$ and increasing the neighborhood with codebook depth.
This is an engineering bandwidth heuristic, not a theoretical equivalence
between perplexity and KNN radius. The temperature is $\tau=1.0$, and every
KNN table is constructed within the corresponding codebook while excluding
the query token.

We use a three-stage EMDC curriculum: clean codec conditions for steps 0--30k,
a linear ramp of each replacement rate from zero to $p_k$ over 30k--40k, and
the target rate $p_k$ through 150k. Codec embeddings are initialized from the
tokenizer codebooks, frozen for the first 5k steps, and then optimized with
FullDiT.

\subsection{Four-Way Classifier-Free Guidance}
\label{sec:cfg}

Codec, lyrics, and caption serve different purposes. Standard classifier-free
guidance compares fully conditioned and unconditional predictions with one
scale \citep{ho2022cfg}. We instead form four predictions from the same model
and flow state:
$f_{klc}$ uses codec, lyrics, and caption;
$f_{kl}$ uses codec and lyrics;
$f_k$ uses codec only; and
$f_u$ is unconditional. The guided field is
\begin{align}
f_{\mathrm{4CFG}}
=\, &f_u
+s_{\mathrm{codec}}(f_k-f_u) \nonumber\\
&+s_{\mathrm{lyr}}(f_{kl}-f_k)
+s_{\mathrm{cap}}(f_{klc}-f_{kl}).
\label{eq:4cfg}
\end{align}
Thus, the three scales control the incremental vector fields contributed by
codec, lyrics, and caption, rather than condition accuracies. When all scales
are one, Equation~\ref{eq:4cfg} reduces to $f_{klc}$, equivalent in output to
joint 2-way CFG with scale one. When lyric and caption scales are equal,
their terms combine, reducing the expression to 3-way codec/joint-text CFG.
When lyrics and captions use different guidance strengths, the full four-way
formulation is required.

\section{Experiments}
\label{sec:experiments}

\subsection{Research Questions and Evaluation Design}
\label{sec:evaluation-design}

Following the method components, we evaluate FullDiT's full-context rendering,
EMDC, 4-CFG, and the complete system in sequence. Matched 1.5B variants support
renderer-level component attribution, whereas evaluations of the 8B renderer
with the upstream planning stack support only end-to-end claims.

\begin{itemize}
\item RQ1: How do full-song context and renderer-side text conditioning
affect rendering? M1--M2a isolates full-song non-causal versus local-window
training context, while M1--M2b isolates renderer-side caption and lyric
conditioning under the same frame-aligned multi-codebook plan.

\item RQ2: Does EMDC improve robustness to predicted codec tokens? M1--M3
isolates error-rate-matched, geometry-guided EMDC while holding architecture,
training budget, full-song context, and renderer-side text conditioning fixed.
Additional cross-codec stress tests appear in the supplement.

\item RQ3: Does full 4-way CFG outperform its 2- and 3-way reductions? With M1
fixed, we sweep seven guidance tuples spanning 2-, 3-, and 4-way CFG.
Automatic screening followed by blind listening selects the best tested
operating point and tests whether separate lyric and caption scales are
preferable to collapsed guidance.

\item RQ4: How competitive is the complete system? We compare it with five
commercial systems on a common set using four automatic evaluation frameworks
and evaluate it on the Artificial Analysis Music with Vocals Leaderboard.

\end{itemize}

Section~\ref{sec:evaluation} specifies the codec sources, comparison systems,
metrics, and listening protocols used to answer these four questions.

\subsection{Experimental Setup}
\label{sec:setup}

\paragraph{Scope and Data}
The complete system uses an 8B FullDiT for end-to-end results. All
component claims use the matched 1.5B variants M1, M2a, M2b, and M3 defined
in Table~\ref{tab:controlled}. The controlled models are trained on the same
amount of internal music data. Audio is read as 48-kHz stereo and encoded by
a fixed LeVo VAE
\citep{lei2025levo} into 25-Hz, 64-dimensional acoustic latents. Offline RVQ
tokens from the semantic-aware tokenizer use the same frame rate.

\paragraph{Architecture}
The 1.5B FullDiT contains 28 Transformer layers, hidden width 1,536, 12 query
heads, two key--value heads, and head dimension 128. Each of the eight RVQ
streams uses a 1,024-entry, 32-dimensional embedding table. Captions are
compact JSON strings encoded by a frozen Qwen3-Embedding-0.6B model
\citep{zhang2025qwen3embedding} and projected to the model width. Lyrics use a
shared token embedding followed by an eight-layer encoder. The combined text
limit is 2,048 tokens, with at most 1,024 each for caption and lyrics.

\paragraph{Training}
Flow matching samples $\sigma$ from a logit-normal distribution with shift
1.0. Training
uses three independent condition masks for codec, caption, and lyrics. For M1,
M2a, and M3, each condition has a dropout probability of 0.1.
M2b replaces the renderer-side caption and lyric conditions with the learned
null condition.
The training-time context, text, and EMDC settings for each model are
summarized in Table~\ref{tab:controlled}.

We use AdamW with peak learning rate $10^{-4}$,
$(\beta_1,\beta_2)=(0.8,0.99)$, weight decay 0.1, and gradient clipping at
1.0. A 4k-step linear warm-up precedes cosine decay. Models train in bfloat16
for 150k steps with global batch size 384. The random seed is 666.

\paragraph{Inference}
All controlled inference uses 50-step UniPC \citep{zhao2023unipc} with
sampling shift 3.0. The 4-CFG study evaluates the seven tuples specified in
the evaluation protocol below. To keep the remaining comparisons controlled,
they use the same listener-preferred operating point
$(s_{\mathrm{codec}},s_{\mathrm{lyr}},s_{\mathrm{cap}})=(1,2,1)$ and keep
all other sampling parameters fixed.

\begin{table*}[t]
\centering
{
\small
\setlength{\tabcolsep}{0.6mm}
\begin{tabular*}{\textwidth}{@{\extracolsep{\fill}}lccccccccccc@{}}
\toprule
\multicolumn{4}{c}{\textbf{Configuration}} &
\multicolumn{3}{c}{\textbf{Clean GT codec}} &
\multicolumn{3}{c}{\textbf{Synthetic corruption}} &
\multicolumn{2}{c}{\textbf{LM-generated codec}} \\
\cmidrule(lr){1-4}\cmidrule(lr){5-7}\cmidrule(lr){8-10}
\cmidrule(lr){11-12}
\textbf{Model} & \textbf{EMDC} &
\textbf{Audio ctx.} &
\textbf{Caption + lyrics} &
\textbf{ViSQOL} &
\textbf{Log-mel L1} &
\textbf{MR-STFT} &
\textbf{ViSQOL} &
\textbf{Log-mel L1} &
\textbf{MR-STFT} &
\textbf{PQ} &
\textbf{Preference} \\
& & & &
$\uparrow$ & $\downarrow$ & $\downarrow$ &
$\uparrow$ & $\downarrow$ & $\downarrow$ &
$\uparrow$ & $\uparrow$ \\
\midrule
\textbf{M1} & \ding{51} & Full & \ding{51} &
3.3388 & 1.1363 & \textbf{1.7806} &
\textbf{3.2036} & \textbf{1.1842} & \textbf{1.7943} &
\textbf{8.213} & Ref. \\
\textbf{M2a} & \ding{51} & 30 s & \ding{51} &
2.8863 & 1.5381 & 1.8707 &
2.6868 & 2.0034 & 2.0190 &
6.011 & 0.0\% \\
\textbf{M2b} & \ding{51} & Full & \ding{55} &
3.3021 & 1.1865 & 1.8253 &
3.1838 & 1.2069 & 1.8327 &
8.135 & 0.0\% \\
\textbf{M3} & \ding{55} & Full & \ding{51} &
\textbf{3.4578} & \textbf{0.9248} & 1.7896 &
2.4342 & 2.2372 & 2.3346 &
8.122 & 30.3\% \\
\bottomrule
\end{tabular*}
}
\caption{Matched 1.5B FullDiT configurations and controlled results across
three codec-token conditions. \ding{51}/\ding{55} denote enabled/disabled
settings; preference is measured against M1. Arrows indicate the better
direction, and bold marks the best value in each metric column.}
\label{tab:controlled}
\end{table*}

\subsection{Evaluation Protocols and Metrics}
\label{sec:evaluation}

\subsubsection{FullDiT Component Evaluation (RQ1--RQ2)}

The matched 1.5B variants are evaluated under three codec-token conditions:
\emph{Clean GT codec}, extracted from held-out target audio;
\emph{Synthetic corruption}, constructed by perturbing the corresponding
clean tokens; and \emph{LM-generated codec}, predicted by the upstream
hierarchical autoregressive language model rather than extracted from
reference audio by the tokenizer.

The clean and synthetic-corrupted conditions use the same $N=200$ held-out
songs with paired acoustic targets. We report ViSQOL
\citep{chinen2020visqol}, log-mel L1 distance, and multi-resolution STFT
distance \citep{yamamoto2020parallelwavegan}; higher ViSQOL and lower distances
are better. Every controlled model
receives exactly the same synthetic-corrupted token sequences. Their
replacement rates come from teacher-forced top-1 errors, while replacement
identities come from M1's step-150k cosine-KNN tables. 

For the LM-generated codec condition, we render $N=200$ songs from one fixed
cache of predicted token sequences. Because these outputs do not have paired
acoustic targets, we assess rendering quality using the reference-free
Audiobox-Aesthetics Production Quality (PQ) score
\citep{tjandra2025audioboxaesthetics} and blind human preference judgments.
PQ is computed on all $N=200$ songs. For human evaluation, blind pairwise
comparisons of M1 against M2a, M2b, and M3 use a common set of $N=100$ songs
and five expert judges with music backgrounds. Each judge chooses audio A,
audio B, or tie without knowing model identity. After unblinding, we report
song-level preference after excluding per-song ties:
$\mathrm{wins}/(\mathrm{wins}+\mathrm{losses})$. Clean GT codec and synthetic
corruption use paired automatic metrics only.

\subsubsection{Four-Way CFG Evaluation (RQ3)}

Fixing M1, we score seven tuples---$(1,1,1)$ and the six obtained by changing
one scale to 0.5 or 2---with Audiobox PQ on $N=200$. The top three enter a blind
three-way forced-choice test on $N=100$: five expert judges each select one
candidate per song, with no ties. We report song-weighted selection shares,
comparing the automatic optimum with listeners' choice.

\subsubsection{Complete-System Evaluation (RQ4)}

We compare the complete system with Suno V5.5 \citep{suno2026v55}, Suno V5
\citep{suno2025v5}, Mureka V8 \citep{mureka2026v8}, Lyria 3 Pro
\citep{google2026lyria3pro}, and MiniMax Music 2.6
\citep{minimax2026music26}. For each case, every system receives the same
user-provided caption and lyrics. All systems are evaluated on the same $N=500$
vocal-music set, spanning eight major genres and balanced across Chinese,
English, Japanese, Korean, and Spanish. SongBench \citep{wu2026songbench}
reports Melody, Arrangement, Musicality, Vocal, Instrument, Mixing, and
Structure. SongEval \citep{yao2025songeval} reports overall coherence, overall
musicality, memorability, structural clarity, and vocal breathing/phrasing
naturalness. Audiobox-Aesthetics reports Content Enjoyment, Content
Usefulness, Production Complexity, and Production Quality. CMI-RM
\citep{ma2026cmirewardbench} reports Alignment and Musicality.

The complete system also participates in the Artificial Analysis Music with
Vocals Leaderboard \citep{artificialanalysis2026vocals}, a public website for
blind human preference evaluation of music generation systems. We report an archived, dated snapshot together with
its evaluation count, Elo estimate, confidence interval, and rank interval.

\begin{table*}[t]
\centering
{
\small
\setlength{\tabcolsep}{1mm}
\begin{tabular*}{0.98\textwidth}{@{\extracolsep{\fill}}lccccccc@{}}
\toprule
  \textbf{Evaluator} & \textbf{Metric} & \textbf{Ours} &
  \textbf{Suno 5.5} & \textbf{Suno 5} & \textbf{Mureka 8} &
  \textbf{Lyria 3 Pro} & \textbf{MiniMax 2.6} \\
\midrule
\textbf{SongBench} & Melody & \textbf{7.2001} & 6.6939 & 6.9202 & 7.0660 & 7.0377 & 6.7245 \\
 & Arrangement & \textbf{7.3879} & 6.8406 & 7.0805 & 7.2601 & 7.1511 & 6.8167 \\
 & Musicality & \textbf{6.3678} & 5.8297 & 6.0096 & 6.2559 & 6.1810 & 5.8823 \\
 & Vocal & 7.6234 & 7.0981 & 7.3160 & \textbf{7.6248} & 7.4560 & 7.2513 \\
 & Instrument & \textbf{7.3517} & 7.0080 & 7.1487 & 7.2585 & 7.0583 & 6.8458 \\
 & Mixing & \textbf{7.3047} & 7.0332 & 7.1101 & 7.1433 & 7.0093 & 6.7397 \\
 & Structure & 6.9650 & 6.6158 & 6.7363 & 7.0237 & \textbf{7.0565} & 6.5637 \\
\midrule
\textbf{SongEval} & Coherence & \textbf{4.5094} & 4.2905 & 4.3278 & 4.3870 & 4.4673 & 4.2772 \\
 & Musicality & \textbf{4.4098} & 4.1547 & 4.1970 & 4.2875 & 4.3466 & 4.1781 \\
 & Memorability & \textbf{4.4663} & 4.2123 & 4.2473 & 4.3680 & 4.4153 & 4.2180 \\
 & Clarity & \textbf{4.3838} & 4.1246 & 4.1741 & 4.2431 & 4.3099 & 4.1448 \\
 & Naturalness & \textbf{4.2600} & 4.0046 & 4.0679 & 4.1767 & 4.2021 & 3.9919 \\
\midrule
\textbf{Audiobox} & Content Enjoyment & \textbf{7.7089} & 7.5289 & 7.5507 & 7.5921 & 7.6950 & 7.6844 \\
 & Content Usefulness & \textbf{7.9989} & 7.9535 & 7.8020 & 7.8634 & 7.8919 & 7.9578 \\
 & Prod. Complexity & \textbf{6.8774} & 6.6486 & 6.8118 & 6.8768 & 6.7561 & 6.7514 \\
 & Production Quality & \textbf{8.2986} & 8.2083 & 8.1411 & 8.1098 & 8.2780 & 8.2923 \\
\midrule
\textbf{CMI-RM} & Alignment & 2.1786 & 1.9216 & 1.9119 & \textbf{2.3089} & 2.0377 & 2.1546 \\
 & Musicality & \textbf{2.7611} & 2.3665 & 2.3756 & 2.7590 & 2.5280 & 2.6661 \\
\bottomrule
\end{tabular*}
}
\caption{Automatic comparison on the common multilingual vocal-music
set. Bold marks the best point estimate for each metric.}
\label{tab:system}
\end{table*}

\section{Results and Analysis}
\label{sec:results}

\subsection{Controlled FullDiT Ablations}
\label{sec:ablations}

All models in Table~\ref{tab:controlled} use the selected 4-CFG tuple
$(1,2,1)$. Clean and synthetic tests share held-out targets; the
language-model study fixes the upstream cache, caption/lyrics, length, seed,
and sampling recipe.

\subsubsection{Frame-Aligned Codec Conditioning and Full-Song Context}

All four controlled models receive the same eight frame-aligned codec streams;
the matched M1--M2a comparison isolates context span as the only varying
factor. M1 and M2a both use EMDC, renderer-side text conditioning, and the same
architecture, but M2a is restricted to a local 30-second context window during
training. It
underperforms M1 on every clean and corrupted metric, and
its language-model-conditioned PQ drops from 8.213 to 6.011. M2a wins 0.0\%
of non-tied songs against M1. These results show that local training windows
are inadequate for the renderer even when the upstream token plan is held
fixed; non-causal full-song training context is central to acoustic
realization.

M2b replaces the renderer-side caption and lyric conditions with the learned
null condition
while retaining M1's full-song context, EMDC, and architecture. M1 is better
on all six paired objective metrics, and its language-model-conditioned PQ is
8.213 rather than 8.135. M2b wins 0.0\% of non-tied songs against M1. Codec
tokens remain the dominant local plan, but direct text access provides
complementary semantic and vocal information that is not fully transmitted
through the finite-rate token sequence.

\subsubsection{Error-Matched Distractor Conditioning}

M1 and M3 have identical 1.5B architecture, full-song context, renderer-side
text conditioning, and 150k-step budget; only EMDC differs. EMDC's advantage
is concentrated at the imperfect codec interface. Under synthetic corruption,
M1 raises
ViSQOL from
2.4342 to 3.2036, reduces log-mel L1 from 2.2372 to 1.1842, and reduces
MR-STFT from 2.3346 to 1.7943. These three complementary measures agree in
both direction and magnitude. With fixed language-model tokens, PQ also rises
from 8.122 to 8.213. M1 wins 69.7\% of non-tied songs against M3. EMDC
therefore exchanges some
clean-condition reconstruction for substantially stronger recovery from
synthetic token errors and better rendering from the deployed interface.

\subsection{Four-Way CFG Analysis}
\label{sec:cfg-results}

  \begin{figure*}[!t]
      \centering
      \includegraphics[width=0.9\textwidth]
      {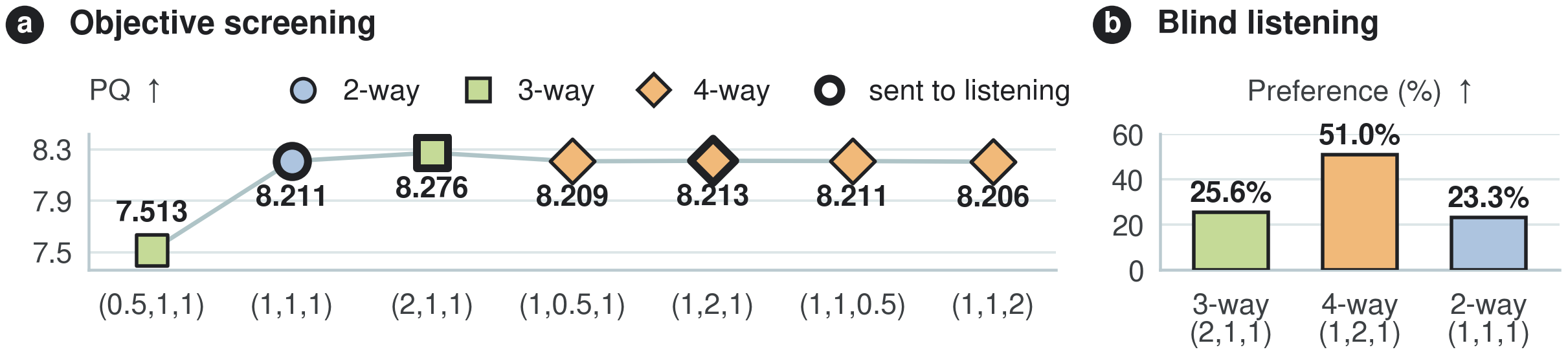}
      \caption{Four-way CFG analysis: (a) Audiobox PQ screening
      of seven guidance
      tuples; (b) song-weighted selection shares from the blind three-way
      forced-choice evaluation of the three selected candidates.}
      \label{fig:cfg}
  \end{figure*}

\begin{figure*}[!t]
    \centering
    \includegraphics[width=0.9\textwidth]{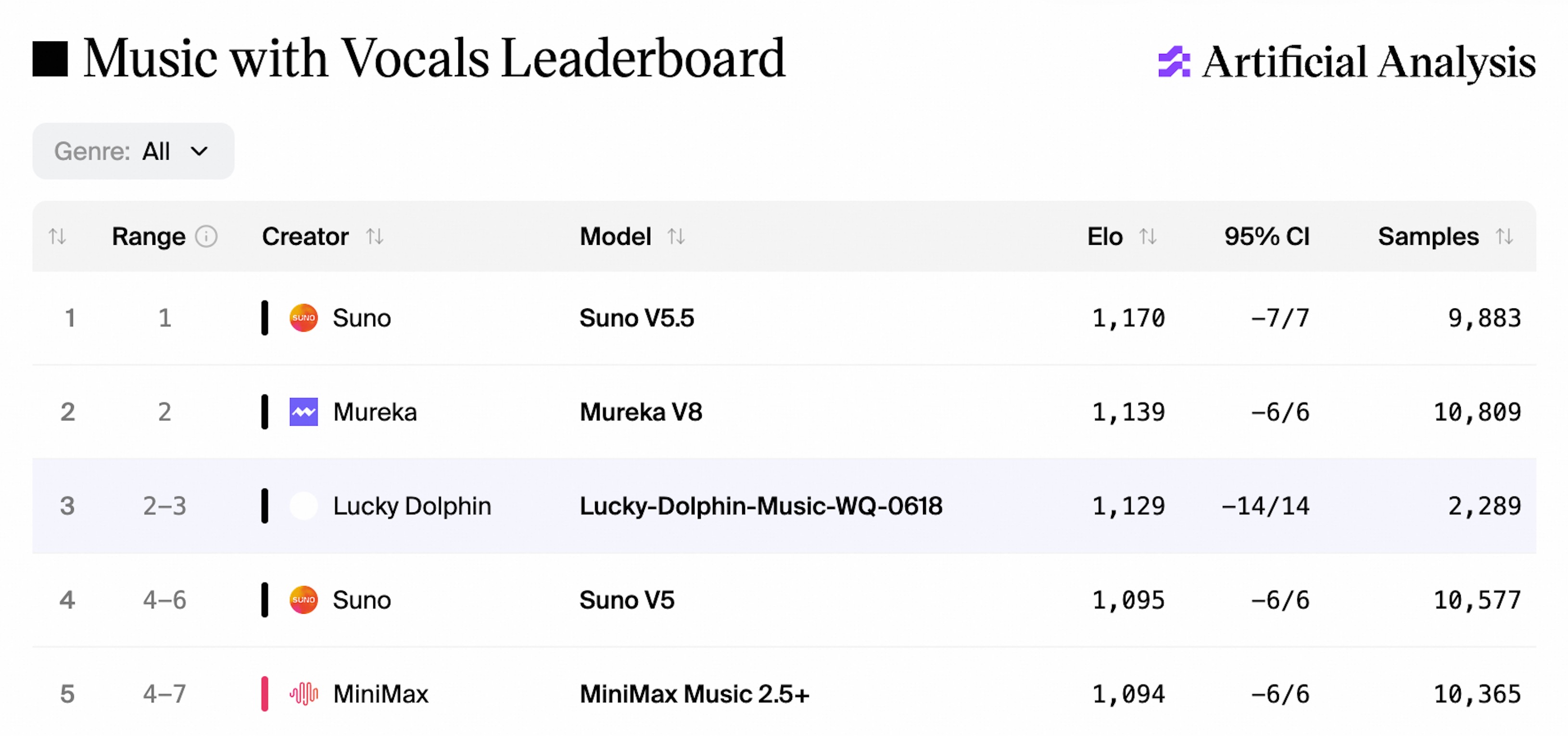}
    \caption{Archived Artificial Analysis Music with Vocals Leaderboard
    snapshot used in this paper. The highlighted Lucky Dolphin entry, listed
    as Lucky-Dolphin-Music-WQ-0618, corresponds to our complete system.}
    \label{fig:aa}
\end{figure*}

We fix M1 and denote each guidance tuple as
$(s_{\mathrm{codec}},s_{\mathrm{lyr}},s_{\mathrm{cap}})$. Starting from
$(1,1,1)$, independently setting one scale to 0.5 or 2 yields seven
configurations. The anchor reduces to fully conditioned output;
$(0.5,1,1)$ and $(2,1,1)$ reduce to 3-way codec/joint-text CFG; the remaining
four require distinct lyric and caption increments.

As Figure~\ref{fig:cfg} shows, we submit the top three settings by automatic
PQ---$(2,1,1)$, $(1,2,1)$, and $(1,1,1)$---to blind listening. The human
three-way forced-choice evaluation selects $(1,2,1)$ with 51.0\% of the
song-weighted selections; we use this tuple for the controlled comparisons in
Table~\ref{tab:controlled}.

\subsection{Complete-System Results (RQ4)}
\label{sec:system-results}

\subsubsection{Common-Set Automatic Evaluation}

Table~\ref{tab:system} evaluates the production system, including upstream
codec planning and the 8B renderer. Given user-provided captions and lyrics,
our system outperforms the five commercial systems on 15 of 18 metrics:
five of seven SongBench dimensions, all five SongEval dimensions, all four
Audiobox dimensions, and CMI-RM Musicality.

\subsubsection{External Public Blind Evaluation}

Our complete system appears as the highlighted Lucky Dolphin entry, listed as
Lucky-Dolphin-Music-WQ-0618, in the archived snapshot in
Figure~\ref{fig:aa}. It receives 1,129 Elo from 2,289 samples, with
point-estimate rank 3, rank interval 2--3, and a 95\% Elo interval of
$\pm14$. This external result measures end-to-end caption-and-lyrics-to-song
preference and supports the complete system's competitiveness.

\section{Conclusion}
\label{sec:conclusion}

We identify codec-interface exposure bias in hybrid music generation: an
acoustic renderer trained on perfectly paired codec conditions is deployed on
imperfect predictions from an upstream token language model. FullDiT responds
by combining frame-aligned multi-codebook codec embeddings, independently
encoded captions and lyrics, and non-causal full-song context in one
conditional flow generator. EMDC adds error-rate-matched, geometry-guided
independent substitutions during training, and 4-CFG separates codec, lyric,
and caption increments at inference time. Controlled experiments expose a trade-off. EMDC produces
large, consistent gains across ViSQOL, log-mel, and MR-STFT under synthetic
corruption, alongside higher production quality and listener preference with
fixed language-model tokens. Full-song context and renderer-side text
conditioning both contribute under matched ablation, while the 4-CFG study
selects $(1,2,1)$ as
the listener-preferred guidance tuple. Together, the
component and complete-system evaluations show that our method preserves the
planning benefits of discrete tokens while improving the robustness of
acoustic realization.

\bibliography{draft}

\begin{thebibliography}{38}
\providecommand{\natexlab}[1]{#1}

\bibitem[{Agostinelli et~al.(2023)Agostinelli, Denk, Borsos, Engel, Verzetti,
  Caillon, Huang, Jansen, Roberts, Tagliasacchi, Sharifi, Zeghidour, and
  Frank}]{agostinelli2023musiclm}
Agostinelli, A.; Denk, T.~I.; Borsos, Z.; Engel, J.; Verzetti, M.; Caillon, A.;
  Huang, Q.; Jansen, A.; Roberts, A.; Tagliasacchi, M.; Sharifi, M.; Zeghidour,
  N.; and Frank, C. 2023.
\newblock {MusicLM}: Generating Music From Text.
\newblock \emph{arXiv preprint arXiv:2301.11325}.

\bibitem[{{Artificial Analysis}(2026)}]{artificialanalysis2026vocals}
{Artificial Analysis}. 2026.
\newblock Music with Vocals Leaderboard.
\newblock \url{https://artificialanalysis.ai/music/leaderboard/vocals}.
\newblock Accessed 2026-07-20; the manuscript reports a dated archived
  snapshot.

\bibitem[{Bai et~al.(2024)Bai, Chen, Chen, Chen, Deng, Dong, Hantrakul, Hao,
  Huang, Huang, Jia, La, Le, Li, Li, Li, Li, Liu, Lu, Lu, Shaw, Spijkervet,
  Sun, Wang, Wang, Wang, Wang, Xu, Yang, Yao, Zhang, Zhang, Zhang, Zhao, Zhao,
  Zhong, Zhou, and Zou}]{bai2024seedmusic}
Bai, Y.; Chen, H.; Chen, J.; Chen, Z.; Deng, Y.; Dong, X.; Hantrakul, L.; Hao,
  W.; Huang, Q.; Huang, Z.; Jia, D.; La, F.; Le, D.; Li, B.; Li, C.; Li, H.;
  Li, X.; Liu, S.; Lu, W.-T.; Lu, Y.; Shaw, A.; Spijkervet, J.; Sun, Y.; Wang,
  B.; Wang, J.-C.; Wang, Y.; Wang, Y.; Xu, L.; Yang, Y.; Yao, C.; Zhang, S.;
  Zhang, Y.; Zhang, Y.; Zhao, H.; Zhao, Z.; Zhong, D.; Zhou, S.; and Zou, P.
  2024.
\newblock {Seed-Music}: A Unified Framework for High Quality and Controlled
  Music Generation.
\newblock \emph{arXiv preprint arXiv:2409.09214}.

\bibitem[{Bengio et~al.(2015)Bengio, Vinyals, Jaitly, and
  Shazeer}]{bengio2015scheduledsampling}
Bengio, S.; Vinyals, O.; Jaitly, N.; and Shazeer, N. 2015.
\newblock Scheduled Sampling for Sequence Prediction with Recurrent Neural
  Networks.
\newblock In \emph{Advances in Neural Information Processing Systems},
  volume~28. Curran Associates, Inc.

\bibitem[{Chinen et~al.(2020)Chinen, Lim, Skoglund, Gureev, O'Gorman, and
  Hines}]{chinen2020visqol}
Chinen, M.; Lim, F. S.~C.; Skoglund, J.; Gureev, N.; O'Gorman, F.; and Hines,
  A. 2020.
\newblock {ViSQOL v3}: An Open Source Production Ready Objective Speech and
  Audio Metric.
\newblock In \emph{2020 Twelfth International Conference on Quality of
  Multimedia Experience}, 1--6.

\bibitem[{Chiu et~al.(2022)Chiu, Qin, Zhang, Yu, and Wu}]{chiu2022bestrq}
Chiu, C.-C.; Qin, J.; Zhang, Y.; Yu, J.; and Wu, Y. 2022.
\newblock Self-Supervised Learning with Random-Projection Quantizer for Speech
  Recognition.
\newblock In \emph{Proceedings of the 39th International Conference on Machine
  Learning}, volume 162 of \emph{Proceedings of Machine Learning Research},
  3915--3924. PMLR.

\bibitem[{Copet et~al.(2023)Copet, Kreuk, Gat, Remez, Kant, Synnaeve, Adi, and
  D{\'e}fossez}]{copet2023musicgen}
Copet, J.; Kreuk, F.; Gat, I.; Remez, T.; Kant, D.; Synnaeve, G.; Adi, Y.; and
  D{\'e}fossez, A. 2023.
\newblock Simple and Controllable Music Generation.
\newblock In \emph{Advances in Neural Information Processing Systems},
  volume~36, 47704--47720. Curran Associates, Inc.

\bibitem[{D{\'e}fossez et~al.(2023)D{\'e}fossez, Copet, Synnaeve, and
  Adi}]{defossez2023encodec}
D{\'e}fossez, A.; Copet, J.; Synnaeve, G.; and Adi, Y. 2023.
\newblock High Fidelity Neural Audio Compression.
\newblock \emph{Transactions on Machine Learning Research}.

\bibitem[{Dhariwal et~al.(2020)Dhariwal, Jun, Payne, Kim, Radford, and
  Sutskever}]{dhariwal2020jukebox}
Dhariwal, P.; Jun, H.; Payne, C.; Kim, J.~W.; Radford, A.; and Sutskever, I.
  2020.
\newblock {Jukebox}: A Generative Model for Music.
\newblock \emph{arXiv preprint arXiv:2005.00341}.

\bibitem[{Evans et~al.(2024)Evans, Carr, Taylor, Hawley, and
  Pons}]{evans2024stableaudio}
Evans, Z.; Carr, C.; Taylor, J.; Hawley, S.~H.; and Pons, J. 2024.
\newblock Fast Timing-Conditioned Latent Audio Diffusion.
\newblock In \emph{Proceedings of the 41st International Conference on Machine
  Learning}, volume 235 of \emph{Proceedings of Machine Learning Research},
  12652--12665. PMLR.

\bibitem[{Gong et~al.(2025)Gong, Zhao, Wang, Xu, and Guo}]{gong2025acestep}
Gong, J.; Zhao, S.; Wang, S.; Xu, S.; and Guo, J. 2025.
\newblock {ACE-Step}: A Step Towards Music Generation Foundation Model.
\newblock \emph{arXiv preprint arXiv:2506.00045}.

\bibitem[{{Google}(2026)}]{google2026lyria3pro}
{Google}. 2026.
\newblock {Lyria 3} Expands to More Google Products and Adds More Features.
\newblock
  \url{https://blog.google/innovation-and-ai/technology/ai/lyria-3-pro/}.
\newblock Product release, accessed 2026-07-20.

\bibitem[{Gulati et~al.(2020)Gulati, Qin, Chiu, Parmar, Zhang, Yu, Han, Wang,
  Zhang, Wu, and Pang}]{gulati2020conformer}
Gulati, A.; Qin, J.; Chiu, C.-C.; Parmar, N.; Zhang, Y.; Yu, J.; Han, W.; Wang,
  S.; Zhang, Z.; Wu, Y.; and Pang, R. 2020.
\newblock Conformer: Convolution-Augmented Transformer for Speech Recognition.
\newblock In \emph{Interspeech 2020}, 5036--5040.

\bibitem[{Ho et~al.(2022)Ho, Saharia, Chan, Fleet, Norouzi, and
  Salimans}]{ho2022cascaded}
Ho, J.; Saharia, C.; Chan, W.; Fleet, D.~J.; Norouzi, M.; and Salimans, T.
  2022.
\newblock Cascaded Diffusion Models for High Fidelity Image Generation.
\newblock \emph{Journal of Machine Learning Research}, 23(47): 1--33.

\bibitem[{Ho and Salimans(2022)}]{ho2022cfg}
Ho, J.; and Salimans, T. 2022.
\newblock Classifier-Free Diffusion Guidance.
\newblock \emph{arXiv preprint arXiv:2207.12598}.

\bibitem[{Huang et~al.(2023)Huang, Park, Wang, Denk, Ly, Chen, Zhang, Zhang,
  Yu, Frank, Engel, Le, Chan, Chen, and Han}]{huang2023noise2music}
Huang, Q.; Park, D.~S.; Wang, T.; Denk, T.~I.; Ly, A.; Chen, N.; Zhang, Z.;
  Zhang, Z.; Yu, J.; Frank, C.; Engel, J.; Le, Q.~V.; Chan, W.; Chen, Z.; and
  Han, W. 2023.
\newblock {Noise2Music}: Text-Conditioned Music Generation with Diffusion
  Models.
\newblock \emph{arXiv preprint arXiv:2302.03917}.

\bibitem[{Lam et~al.(2023)Lam, Tian, Li, Yin, Feng, Tu, Ji, Xia, Ma, Song,
  Chen, Wang, and Wang}]{lam2023melody}
Lam, M. W.~Y.; Tian, Q.; Li, T.; Yin, Z.; Feng, S.; Tu, M.; Ji, Y.; Xia, R.;
  Ma, M.; Song, X.; Chen, J.; Wang, Y.; and Wang, Y. 2023.
\newblock Efficient Neural Music Generation.
\newblock In \emph{Advances in Neural Information Processing Systems},
  volume~36, 17450--17463.

\bibitem[{Lei et~al.(2025)Lei, Xu, Lin, Zhang, Tan, Chen, Zhang, Yang, Zhu,
  Wang, Wu, and Yu}]{lei2025levo}
Lei, S.; Xu, Y.; Lin, Z.; Zhang, H.; Tan, W.; Chen, H.; Zhang, Y.; Yang, C.;
  Zhu, H.; Wang, S.; Wu, Z.; and Yu, D. 2025.
\newblock {LeVo}: High-Quality Song Generation with Multi-Preference Alignment.
\newblock In \emph{Advances in Neural Information Processing Systems},
  volume~38.

\bibitem[{Lipman et~al.(2023)Lipman, Chen, Ben-Hamu, Nickel, and
  Le}]{lipman2023flowmatching}
Lipman, Y.; Chen, R. T.~Q.; Ben-Hamu, H.; Nickel, M.; and Le, M. 2023.
\newblock Flow Matching for Generative Modeling.
\newblock In \emph{The Eleventh International Conference on Learning
  Representations}.

\bibitem[{Liu et~al.(2024)Liu, Wang, Gong, and Glass}]{liu2024codecresynthesis}
Liu, A.~H.; Wang, Q.; Gong, Y.; and Glass, J. 2024.
\newblock A Closer Look at Neural Codec Resynthesis: Bridging the Gap between
  Codec and Waveform Generation.
\newblock \emph{arXiv preprint arXiv:2410.22448}.
\newblock NeurIPS 2024 Audio Imagination Workshop.

\bibitem[{Ma et~al.(2026)Ma, Xia, Gao, Chen, Ye, Yang, Chang, Ding, Li, Yuan,
  Dixon, and Benetos}]{ma2026cmirewardbench}
Ma, Y.; Xia, H.; Gao, H.; Chen, W.; Ye, Y.; Yang, Y.; Chang, S.; Ding, M.; Li,
  Y.; Yuan, R.; Dixon, S.; and Benetos, E. 2026.
\newblock {CMI-RewardBench}: Evaluating Music Reward Models with Compositional
  Multimodal Instruction.
\newblock \emph{arXiv preprint arXiv:2603.00610}.
\newblock Accepted by ICML 2026.

\bibitem[{{MiniMax}(2026)}]{minimax2026music26}
{MiniMax}. 2026.
\newblock {MiniMax Music 2.6}: Four Stories We Want to Tell.
\newblock \url{https://www.minimax.io/news/music-26}.
\newblock Product release, accessed 2026-07-20.

\bibitem[{{Mureka}(2026)}]{mureka2026v8}
{Mureka}. 2026.
\newblock Mureka {API} Platform Changelog.
\newblock \url{https://platform.mureka.ai/docs/en/changelog.html}.
\newblock Documents the enhanced mureka-8 release; accessed 2026-07-20.

\bibitem[{Ning et~al.(2025)Ning, Chen, Jiang, Hao, Ma, Wang, Yao, and
  Xie}]{ning2025diffrhythm}
Ning, Z.; Chen, H.; Jiang, Y.; Hao, C.; Ma, G.; Wang, S.; Yao, J.; and Xie, L.
  2025.
\newblock {DiffRhythm}: Blazingly Fast and Embarrassingly Simple End-to-End
  Full-Length Song Generation with Latent Diffusion.
\newblock \emph{arXiv preprint arXiv:2503.01183}.

\bibitem[{Peebles and Xie(2023)}]{peebles2023dit}
Peebles, W.; and Xie, S. 2023.
\newblock Scalable Diffusion Models with Transformers.
\newblock In \emph{Proceedings of the IEEE/CVF International Conference on
  Computer Vision}, 4195--4205.

\bibitem[{San~Roman et~al.(2023)San~Roman, Adi, Deleforge, Serizel, Synnaeve,
  and D{\'e}fossez}]{sanroman2023multibanddiffusion}
San~Roman, R.; Adi, Y.; Deleforge, A.; Serizel, R.; Synnaeve, G.; and
  D{\'e}fossez, A. 2023.
\newblock From Discrete Tokens to High-Fidelity Audio Using Multi-Band
  Diffusion.
\newblock In \emph{Advances in Neural Information Processing Systems},
  volume~36.

\bibitem[{Shulman(2026)}]{suno2026v55}
Shulman, M. 2026.
\newblock {Suno v5.5}: More Expressive. More You.
\newblock \url{https://suno.com/blog/v5-5}.
\newblock Product release, accessed 2026-07-20.

\bibitem[{{Suno}(2025)}]{suno2025v5}
{Suno}. 2025.
\newblock Introducing v5.
\newblock
  \url{https://suno.com/release-notes/introducing-v5-the-world-s-best-music-model}.
\newblock Product release note, accessed 2026-07-20.

\bibitem[{Tjandra et~al.(2025)Tjandra, Wu, Guo, Hoffman, Ellis, Vyas, Shi,
  Chen, Le, Zacharov, Wood, Lee, and Hsu}]{tjandra2025audioboxaesthetics}
Tjandra, A.; Wu, Y.-C.; Guo, B.; Hoffman, J.; Ellis, B.; Vyas, A.; Shi, B.;
  Chen, S.; Le, M.; Zacharov, N.; Wood, C.; Lee, A.; and Hsu, W.-N. 2025.
\newblock {Meta Audiobox Aesthetics}: Unified Automatic Quality Assessment for
  Speech, Music, and Sound.
\newblock \emph{arXiv preprint arXiv:2502.05139}.

\bibitem[{Wu et~al.(2026)Wu, Lei, Tan, Li, Wang, Zhang, Zuo, and
  Wu}]{wu2026songbench}
Wu, D.; Lei, S.; Tan, W.; Li, G.; Wang, Y.; Zhang, H.; Zuo, L.; and Wu, Z.
  2026.
\newblock {SongBench}: A Fine-Grained Multi-Aspect Benchmark for Song Quality
  Assessment.
\newblock \emph{arXiv preprint arXiv:2604.25937}.

\bibitem[{Yamamoto, Song, and Kim(2020)}]{yamamoto2020parallelwavegan}
Yamamoto, R.; Song, E.; and Kim, J.-M. 2020.
\newblock {Parallel WaveGAN}: A Fast Waveform Generation Model Based on
  Generative Adversarial Networks with Multi-Resolution Spectrogram.
\newblock In \emph{2020 IEEE International Conference on Acoustics, Speech and
  Signal Processing}, 6199--6203.

\bibitem[{Yang et~al.(2025)Yang, Wang, Chen, Tan, Yu, and
  Li}]{yang2025songbloom}
Yang, C.; Wang, S.; Chen, H.; Tan, W.; Yu, J.; and Li, H. 2025.
\newblock {SongBloom}: Coherent Song Generation via Interleaved Autoregressive
  Sketching and Diffusion Refinement.
\newblock In \emph{Advances in Neural Information Processing Systems},
  volume~38.

\bibitem[{Yang et~al.(2026)Yang, Xie, Yin, Wang, Yi, Zhu, Weng, Xiong, Ma,
  Cong, Liu, Huang, Ru, Huang, Wan, Wang, Yu, Wang, Liang, Zhuang, Wang, Wang,
  Guo, Cao, Ju, Liu, Cao, Weng, and Zou}]{yang2026heartmula}
Yang, D.; Xie, Y.; Yin, Y.; Wang, Z.; Yi, X.; Zhu, G.; Weng, X.; Xiong, Z.; Ma,
  Y.; Cong, D.; Liu, J.; Huang, Z.; Ru, J.; Huang, R.; Wan, H.; Wang, P.; Yu,
  K.; Wang, H.; Liang, L.; Zhuang, X.; Wang, Y.; Wang, D.; Guo, H.; Cao, J.;
  Ju, Z.; Liu, S.; Cao, Y.; Weng, H.; and Zou, Y. 2026.
\newblock {HeartMuLa}: A Family of Open Sourced Music Foundation Models.
\newblock \emph{arXiv preprint arXiv:2601.10547}.

\bibitem[{Yao et~al.(2025)Yao, Ma, Xue, Chen, Hao, Jiang, Liu, Yuan, Xu, Xue,
  Liu, and Xie}]{yao2025songeval}
Yao, J.; Ma, G.; Xue, H.; Chen, H.; Hao, C.; Jiang, Y.; Liu, H.; Yuan, R.; Xu,
  J.; Xue, W.; Liu, H.; and Xie, L. 2025.
\newblock {SongEval}: A Benchmark Dataset for Song Aesthetics Evaluation.
\newblock \emph{arXiv preprint arXiv:2505.10793}.

\bibitem[{Zeghidour et~al.(2022)Zeghidour, Luebs, Omran, Skoglund, and
  Tagliasacchi}]{zeghidour2022soundstream}
Zeghidour, N.; Luebs, A.; Omran, A.; Skoglund, J.; and Tagliasacchi, M. 2022.
\newblock {SoundStream}: An End-to-End Neural Audio Codec.
\newblock \emph{IEEE/ACM Transactions on Audio, Speech, and Language
  Processing}, 30: 495--507.

\bibitem[{Zhang et~al.(2025{\natexlab{a}})Zhang, Ma, Chen, Wang, Zhao, Pan,
  Wang, Ni, Nguyen, Zhou, Jiang, Tan, Gao, Du, and Ma}]{zhang2025inspiremusic}
Zhang, C.; Ma, Y.; Chen, Q.; Wang, W.; Zhao, S.; Pan, Z.; Wang, H.; Ni, C.;
  Nguyen, T.~H.; Zhou, K.; Jiang, Y.; Tan, C.; Gao, Z.; Du, Z.; and Ma, B.
  2025{\natexlab{a}}.
\newblock {InspireMusic}: Integrating Super Resolution and Large Language Model
  for High-Fidelity Long-Form Music Generation.
\newblock \emph{arXiv preprint arXiv:2503.00084}.

\bibitem[{Zhang et~al.(2025{\natexlab{b}})Zhang, Li, Long, Zhang, Lin, Yang,
  Xie, Yang, Liu, Lin, Huang, and Zhou}]{zhang2025qwen3embedding}
Zhang, Y.; Li, M.; Long, D.; Zhang, X.; Lin, H.; Yang, B.; Xie, P.; Yang, A.;
  Liu, D.; Lin, J.; Huang, F.; and Zhou, J. 2025{\natexlab{b}}.
\newblock {Qwen3 Embedding}: Advancing Text Embedding and Reranking Through
  Foundation Models.
\newblock \emph{arXiv preprint arXiv:2506.05176}.

\bibitem[{Zhao et~al.(2023)Zhao, Bai, Rao, Zhou, and Lu}]{zhao2023unipc}
Zhao, W.; Bai, L.; Rao, Y.; Zhou, J.; and Lu, J. 2023.
\newblock {UniPC}: A Unified Predictor-Corrector Framework for Fast Sampling of
  Diffusion Models.
\newblock In \emph{Advances in Neural Information Processing Systems},
  volume~36, 49842--49869. Curran Associates, Inc.

\end{thebibliography}

\end{document}